\documentclass[namedreferences]{solarphysics}
\usepackage[optionalrh]{spr-sola-addons} 
\usepackage{graphicx}        
\usepackage{amssymb}        
\usepackage{color}           
\usepackage{url}             
 \usepackage{subfig}

\newcommand{\etal}{{\it et al.}}
\newcommand{\todash}{\,--\,}

\begin{document}

\begin{article}

\begin{opening}

\title{Dynamics of Coronal Bright Points as seen by  \textit{ Sun Watcher using Active Pixel System detector and Image
Processing} (SWAP), \textit{Atmospheric
Imaging Assembly}(AIA), and \textit{Helioseismic and Magnetic Imager}(HMI)}

\author{K.~\surname{Chandrashekhar}$^{1}$\sep 
  S.~\surname{Krishna Prasad}$^{1}$\sep
 D.~\surname{Banerjee}$^{1}$\sep
  B.~\surname{Ravindra}$^{1}$\sep
  Daniel~\surname{B.~Seaton}$^{2}$
       }
\runningauthor{K. Chandrashekhar \etal}
\runningtitle{Dynamics of BPs as seen by SWAP, AIA, and HMI}

   \institute{$^{1}$ Indian Institute of Astrophysics,  Koramangala,  Bangalore--560034, India.
                  email: \url{kcs@iiap.res.in} \\
              $^{2}$ SIDC-Royal Observatory of Belgium, Avenue Circulaire 3, 
1180 Brussels, Belgium \\
             }

\begin{abstract}
                      The \textit{Sun Watcher using Active Pixel system detector and Image Processing}(SWAP) on board the \textit{PRoject for OnBoard Autonomy\todash 2} (PROBA\todash 2) spacecraft provides images of the solar corona in EUV channel centered at 174~\AA. These data, together with \textit{Atmospheric Imaging Assembly} (AIA) and the \textit{Helioseismic and Magnetic Imager} (HMI) on board \textit{Solar Dynamics Observatory} (SDO), are used to study the dynamics of coronal bright points. The  evolution  of the magnetic polarities and  associated changes in morphology are studied using magnetograms and multi-wavelength imaging. The morphology of the bright points seen in low-resolution SWAP images and high-resolution AIA images show different structures, whereas the intensity variations with time show similar trends in both SWAP 174 and AIA 171 channels. We observe that bright points are seen in EUV channels corresponding to a magnetic-flux of the order of $10^{18}$~Mx. We find that there exists a good correlation between total emission from the bright point in several UV\todash EUV channels and total unsigned photospheric magnetic flux above certain thresholds. The bright points also show periodic brightenings and we have attempted to find the oscillation periods in bright points and their connection to magnetic flux changes. The observed periods are generally long (10\todash 25~minutes) and there is an indication that the intensity oscillations may be generated by repeated magnetic reconnection.
\end{abstract}
\end{opening}

\section{Introduction}
     \label{S-Introduction} 

 X-ray bright points (XBPs) are prominent dynamical features in quiet-Sun and coronal hole regions. These features were first seen in rocket X-ray telescope images and were reported as small X-ray emitters with a spatial size of less than 60~arcseconds and lifetime ranging from a few hours to a few days \cite{1973ApJ...185L..47V}. \inlinecite{1974ApJ...189L..93G} determined the properties of XBPs from a study of \textit{Skylab} X-ray images. They found the lifetime of the bright points to be about eight hours in X-ray images. Bright points are also observed in EUV images \cite{1981SoPh...69...77H} where the emission has an average lifetime of approximately 20~hours as reported by \inlinecite{2001SoPh..198..347Z}. \inlinecite{1976SoPh...49...79G} estimated the average number of bright points emerging each day to be about 1500. They have also observed a class of long-lived bright points that exists near the Equator, with an average lifetime of about 36~hours.

Bright points are always found to be associated with small, opposite-polarity poles in photospheric magnetograms, which have typical total flux of $10^{19}--10^{20}$~Mx  \cite{1976SoPh...50..311G}. XBPs are likely signatures of small loops that connect the opposite polarities of some small-scale bipoles. It is estimated that one third of the bright points lie over ephemeral regions, which are newly emerging regions of magnetic flux, whereas the remaining two thirds lie above canceling magnetic features, which consist of opposite polarity fragments that approach one another and disappear \cite{1985AuJPh..38..875H,1993SoPh..144...15W}. This process normally takes place at the network boundaries of super granular cells \cite{1990ApJ...352..333H,2003A&A...403..731M,1983SvAL....9..385E}. 

\inlinecite{2001ApJ...547.1100H} found that only magnetic flux above a threshold of $3\times 10^{18}$~Mx is associated with a noticeable brightening in the EIT Fe~{\sc xii} corona. Emerging bipoles typically appear as a brightening on \textit{Yohkoh/SXT} images for only a small percentage of bipoles and have shorter lifetime in X-rays than in EUV, as was found by \inlinecite{2001SoPh..198..347Z}.  A simple explanation given by \inlinecite{2003A&A...398..775M} is that the appearance at a certain temperature depends strongly on the magnetic energy released and, therefore, on the bipole magnetic-field strength. \inlinecite{2004A&A...418..313U} showed that not only does the coronal emission rise due to an increase in emission area, but also because the BP emits more per unit area as the magnetic flux becomes stronger. Whenever there is enhanced emission it is always associated with enhanced magnetic flux. They also found that there is a one-to-one correlation between the magnetic flux in the bipolarity and the EIT Fe~{\sc xii} coronal emission for the two BPs, both in the growing and decaying phase. The exception is that the brightest moments of the BP's lifetime, when there is strong increase in emission, are not accompanied by a comparable increase in the magnetic flux.

The converging-flux model proposed by \inlinecite{1994ApJ...427..459P} describes how the approach of two opposite polarities creates an X-point that rises into the corona and produces a XBP by coronal reconnection \cite{1994SoPh..151...57P,1998ApJ...507..433L}. The model proposes a three-phase evolution: pre-interaction (approach), interaction, and finally cancellation. Several observational questions are raised concerning the evolution of these features. It would be interesting to check the timing of the disappearance of the emission at different temperatures using new observations.

Another set of important characteristics of BP dynamics are oscillations in intensity. Several studies in EUV and soft X-ray spectral lines have reported a wide range of periodicities \cite{1979SoPh...63..119S,1979SoPh...63..113N,1992PASJ...44L.161S,2011MNRAS.415.1419K,2008A&A...489..741T}. \inlinecite{1979SoPh...63..119S} found that the constituent loops could evolve on a time scale of $\approx$ six minutes. \inlinecite{1981SoPh...69...77H} and \inlinecite{1990ApJ...352..333H}, using \textit{Skylab}, showed that BPs exhibit large variations in the emission of chromospheric, transition region, and coronal lines, and no regular periodicity or obvious correlation between the different temperatures was found. It is claimed that acoustic waves, which leak through the magnetic field lines of the solar atmosphere, can be converted into magnetoacoustic waves in the region where the plasma $\beta$ tends to unity and can reach the upper solar atmosphere \cite{2003ApJ...599..626B,2008AnGeo..26.2983K,2010MNRAS.405.2317S}. It is also reported that various intensity oscillations may also be generated by repeated magnetic reconnection \cite{2004PhDT.........1U,2005A&A...442.1087P,2006A&A...446..327D,2008A&A...489..741T}.

Oscillations and the overall temporal evolution of BPs are especially important considerations for determining BPs' overall contribution to the coronal heating problem. In this article we discuss the evolution and dynamics of EUV bright points.  We study the formation, lifetimes, temporal evolution and association with the photospheric magnetic field of several BPs in SWAP, and AIA images.

\section{Observations and Data Analysis}
\label{S-observations} 
   \subsection{SDO} 
   \label{S-sdo}

  The \textit{Solar Dynamics Observatory} (SDO) is the first mission of NASA's Living With a Star (LWS) Program. It was launched on 11 February 2010, and allows nearly continuous observations of the Sun. The \textit{Atmospheric Imaging Assembly} (AIA) on SDO provides an unprecedented view of the solar corona in multiple EUV wavelengths nearly simultaneously. The spatial and temporal resolutions of AIA are 0.6~arcsec per pixel and 12~seconds respectively. AIA contains four telescopes and provides narrow-band imaging in seven EUV bandpasses which are centered on Fe~{\sc xviii} (94~\AA), Fe~{\sc  viii, xxi} (131~\AA),  Fe~{\sc ix} (171~\AA),  Fe~{\sc xii, xxiv} (193~\AA),  Fe~{\sc  xiv}(211~\AA),  He~{\sc ii}(304~\AA),  and Fe~{\sc xvi} (335~\AA) spectral lines.Further details can be found in \inlinecite{2011SoPh..tmp..172L}. The \textit{Helioseismic and Magnetic Imager} (HMI) on SDO is designed to study the magnetic field at the solar surface. It provides full-disk magnetograms at 6173~\AA\ with resolution of 0.5~arcsec per pixel. The temporal resolution of line-of-sight magnetograms from HMI is 45~seconds.

For this study, we have used full-disk, Level~1.0 AIA and HMI images obtained during a period that ran from 13 February, to 15 February 2011. Level~1.0 data are processed to remove bad pixels, spikes due to radiation effects, and correct the image flat-field. Level~1.0 data are then converted to level~1.5 using {\sf aia\_prep.pro} procedure in {\sf Solar Soft} (SSW). This procedure also adjusts the different filter images to a common 0.6$^{\prime\prime}$ resolution. The BPs we studied here have typical lifetimes of at least 20~hours, so we must also correct for solar rotation, which has been done using the SSW procedure {\sf rot\_xy.pro}.

  \begin{figure}    
   \centerline{\includegraphics[width=1.05\textwidth, clip=]{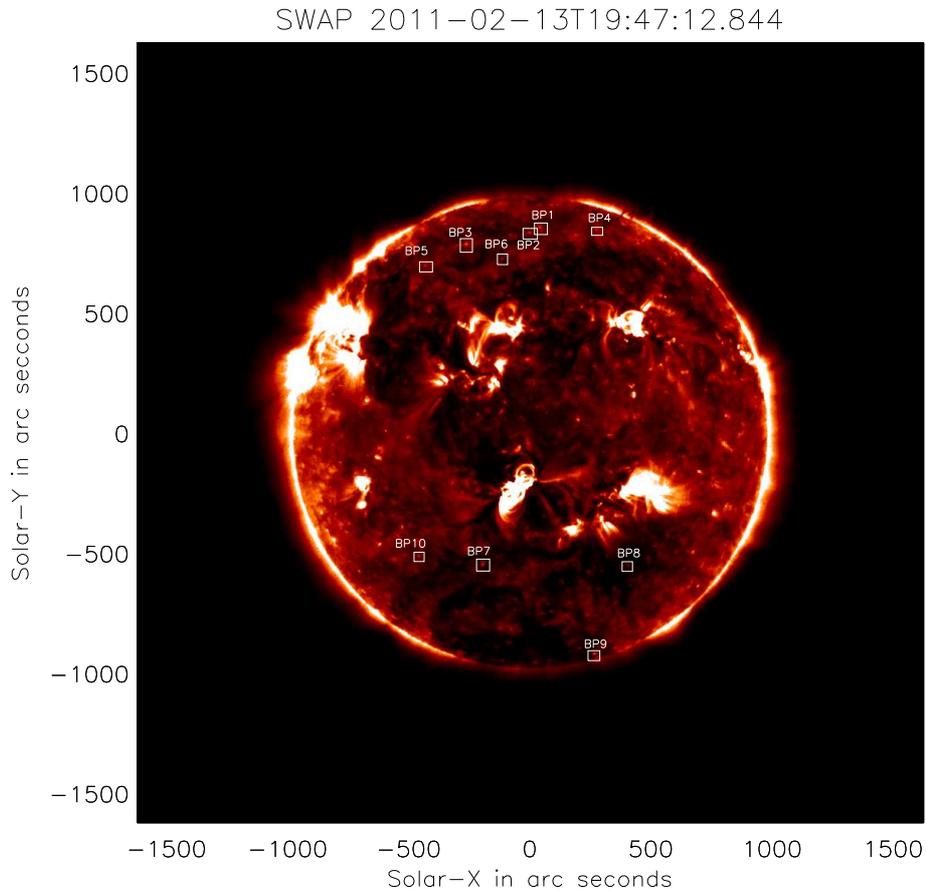}
              }
              \caption{SWAP full-disk image obtained on 13 February 2011, 19:47~UT. The overlaid boxes mark the locations of the BPs studied here.}
   \label{F-swapbps}
   \end{figure}

 \begin{figure}    
   \centerline{\includegraphics[width=1.05\textwidth,
clip=]{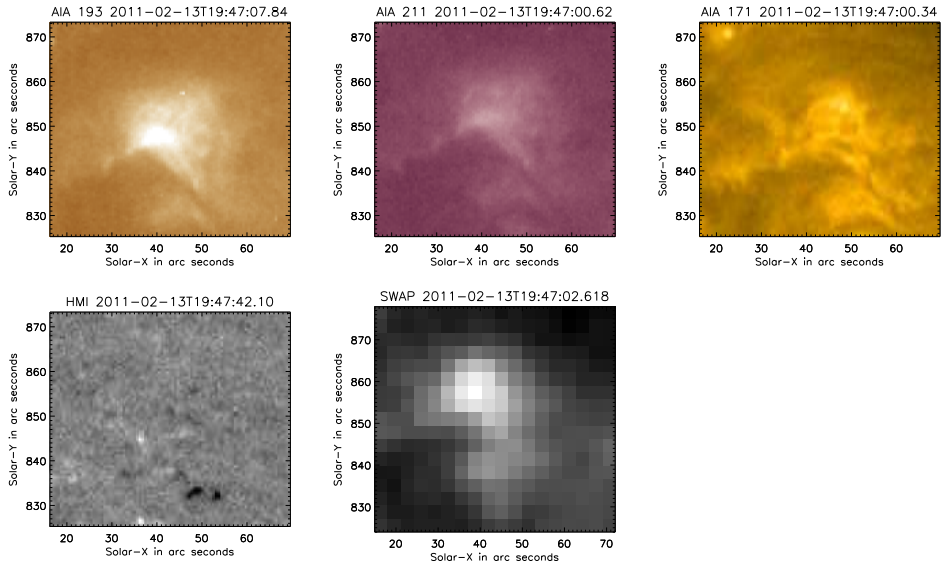}
              }
              \caption{ BP1 as seen in different channels of AIA (top row),  HMI
(bottom left) and SWAP (bottom right). }
   \label{F-swapbps2}
   \end{figure}

\subsection{SWAP} 
\label{S-swap}

In addition to AIA images, we also used data from the \textit{Sun Watcher using Active Pixel System detector and Image Processing} (SWAP) on board the PROBA2 spacecraft \cite{SeatonSWAP2012,HalainSWAP2012,2006AdSpR..38.1807B,2008SoPh..249..147D}. SWAP provides images of the solar corona in an EUV channel centered at 174~\AA\ with a bandpass width of about 15~\AA. SWAP images have a large, square field-of-view that is about 54~arcmin on a side. Pixel size is $\approx 3.16^{\prime\prime}$ and the nominal cadence of the data is roughly 100~seconds. In this study we use calibrated Level-1 data obtained from the PROBA2 Science Center. To produce Level-1 images, raw data from the spacecraft undergoes several processing steps$\colon$  removal of bad and saturated pixels, dark subtraction, cosmic-ray despiking, and an image transformation to that centers the Sun in the frame, rotates solar north to the top of the frame, and corrects plate-scale anisotropy.  Data are also time-normalized based on image integration time.
   
\section{Results} 
      \label{S-results} 
\subsection{Evolution of the Bright Point}

For simplicity, in this section we will focus our attention on one of the typical bright points, identified as BP1 in Figure~\ref{F-swapbps}. Figure~\ref{F-swapbps2} shows a close-up view of BP1 as seen from different instruments.  EUV images in three channels centered at 171~\AA, 193~\AA\,   and 211~\AA\ from AIA are shown in the top row, the corresponding magnetogram is shown in the bottom left and the bottom right image is from SWAP. Previously, when seen by instruments with lower resolution, bright points were believed to be simple loop systems. SWAP images, which have lower resolution than AIA, also can give this impression. However, AIA images show finer structures and multiple connectivity. In fact, this bright point as seen in the AIA image looks like a miniature active region with multiple magnetic poles with several connectivities. BPs no longer can be called a point or simple loop-like structure.  Moreover, depending on the emergence and cancellation of the magnetic polarities, this structure evolves with time. Changes in morphology are very clearly visible in the AIA 193 and 211 channels.

\begin{figure}    
   \centerline{\includegraphics[width=1.0\textwidth, clip=]{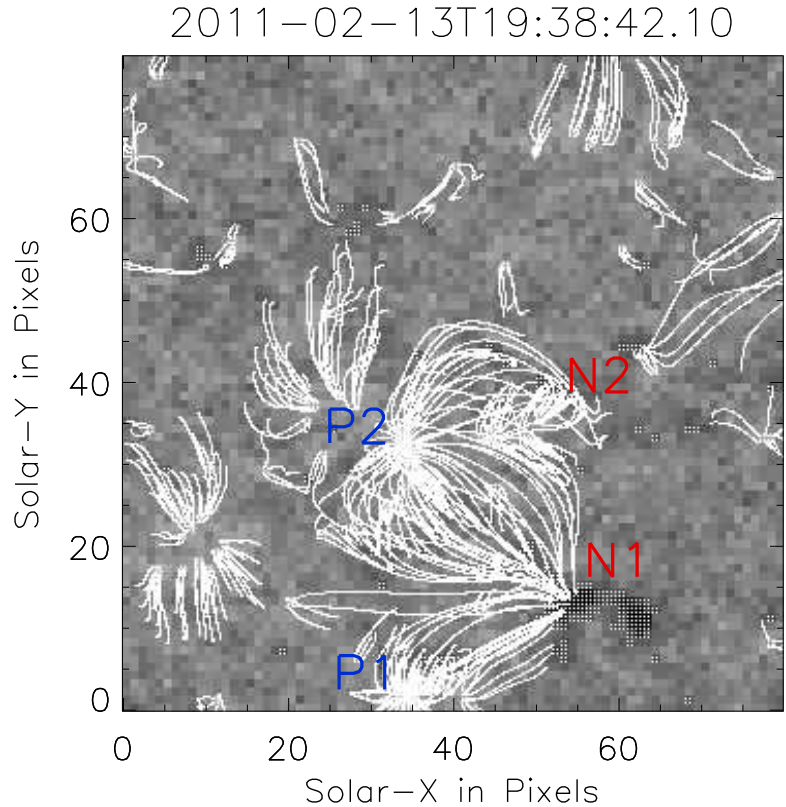}
              }
              \caption{A portion of the HMI line-of-sight magnetogram corresponding to the spatial location of BP1 is shown with the field lines computed using the potential field extrapolation. P1 and P2 indicate the polarities for the first connectivity. P2 and N2 are the polarities for the second connectivity.}
   \label{F-polarities}
   \end{figure}

There are at least three different significant connectivities that we observe in BP1. Figure~\ref{F-polarities} shows the magnetic-field configurations as seen in the HMI magnetograms. The figure also shows field-line connectivity computed using the potential field extrapolation technique (shown as white lines). In the HMI time sequence the polarities that appeared first have been marked as N1\todash P1  and, correspondingly we refer to them as the first connectivity. As the BP evolves, new connectivities form. Subsequently, what we refer to as the second connectivity N2\todash P2, has been indicated as well. Here P1 and N1 are the positive and negative polarities corresponding to the first connectivity, which we see at the location of the bright point at 19:38~UT.  P2 and N2 are the positive and negative polarities for the second connectivity. The white lines connecting different polarities are the potential-field lines drawn using a constant $\alpha$ force-free magnetic field (with $\alpha = 0$) \cite{1981A&A...100..197A,1972SoPh...25..127N}.

\begin{figure}    
   \centerline{\includegraphics[width=0.9\textwidth, clip=]{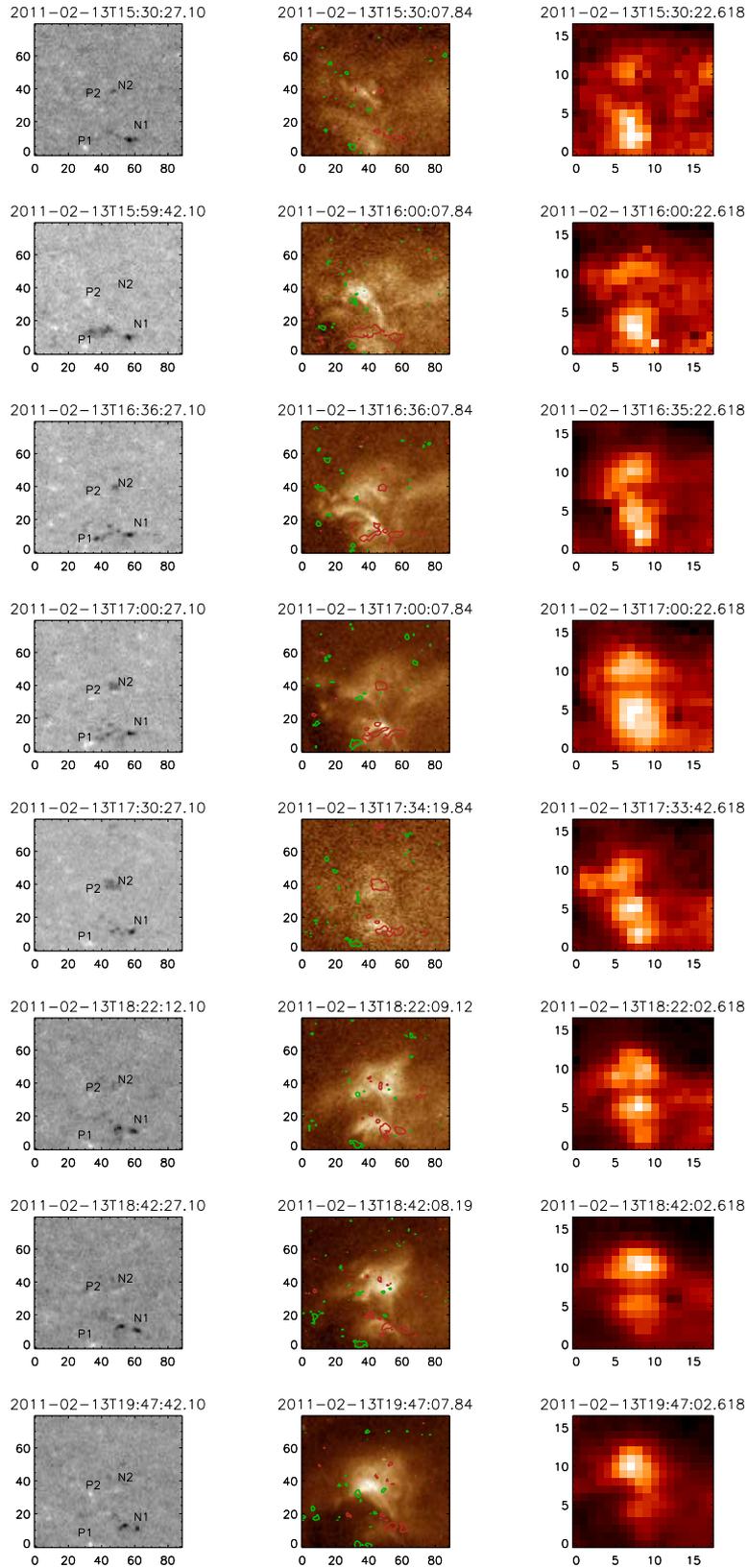}
              }
              \caption{A detailed view of the temporal evolution of BP1. The first column shows the changes in the underlying magnetic field for BP1 from SDO/HMI.  The second column shows the images of BP1 in 193~\AA\ SDO/AIA images. The third column corresponds to the intensity images taken from SWAP 174~\AA~channel. In all the images X and Y axes are in pixel units. }
   \label{F-panel1}
   \end{figure}

From the HMI magnetograms we observe that the two polarities, P1 and N1, are seen beginning at 10:58~UT, but we do not see any enhanced intensity in the loop structure from the AIA channels until around  15:30~UT. However, we observe brightenings in the AIA~171 and 193 channels and SWAP~174 images corresponding to magnetic-flux emergence at the N1 location, during which the area of N1 increased gradually. Similarly, we see a small brightening connecting the locations of P2 and N2. This is shown in Figure~\ref{F-panel1} (first row). Figure~\ref{F-panel1} also shows the temporal evolution of this BP. The first column shows the HMI line-of-sight magnetograms, the second column corresponds to the AIA~193 channel images, and the third column shows the images from SWAP for the corresponding times. The date and time of each image are displayed on top of each snapshot. On AIA images red, and green curves show the contours of $\pm25.0$~G from the HMI magnetograms.

In the first row of Figure~\ref{F-panel1} we observe that in the initial phase of the bright-point evolution (around 15:30~UT)  we hardly see the P2 and N2 polarities. But beginning around 16:00~UT (second row) we see N1 and P2 developing and N2 strength decreasing. Correspondingly, we see connectivity in AIA~193 channel and the SWAP image. Note that here we see loop structures connecting N1 to both P1 and P2 in the AIA~193 channel.  However, in SWAP the N1\todash P1 connectivity appears bright point-like and N2\todash P2 connectivity appears as a diffuse loop, while the N1\todash P2 connectivity is not visible. We attribute this to the lower resolution of SWAP as this connectivity is visible in AIA~171 channel, although not so broad and clear as that in AIA~193.

Around 16:36~UT (third row) at the location of N2 new flux emerges and, correspondingly, we see N2\todash P2 connectivity as a loop structure in both the AIA and SWAP images. A closer look at the third and fourth row of Figure~\ref{F-panel1}reveals that at the location of P1 there is also emergence of new flux, while the negative flux moves towards N1. The area and strength of N1 increases at the same time. At around 17:30~UT (Figure~\ref{F-panel1}, fifth row) we see that the N2 polarity region has grown and connectivity of P2 is now confined to N2 only. As a result of this connectivity we see an enhancement in intensity of AIA lines, but we also observe that there is considerable decrease in N2 strength as time elapses.

At around 18:22 UT, although new flux is still emerging at N2, its strength is decreasing. This could be due to the connectivity with P2. Meanwhile the field strength and size of N1 grows continuously and P1 decreases. As a result of this, at around 19:47~UT we see a clear connectivity between N1 and P2. This change in connectivity from N1\todash P1 and N2\todash P2 to N1\todash P2 is clearly visible in SWAP images. The connectivity changes in different pairs of polarities within BP1 continues for more than 30 hours. We conclude that a combination of flux emergence, annihilation, and perhaps repeated reconnection feed the BP for its entire lifetime until all of the flux within that area vanishes. (A movie of the HMI magnetogram from 13 February 10:58~UT, to 14 February 11:24~UT, and an AIA~193 channel movie from February 13, 15:00~UT to 22:00 UT can be found at \url{ftp://ftp.iiap.res.in/chandra/}.)

 \begin{figure}    
   \centerline{\includegraphics[width=0.95\textwidth,clip=]{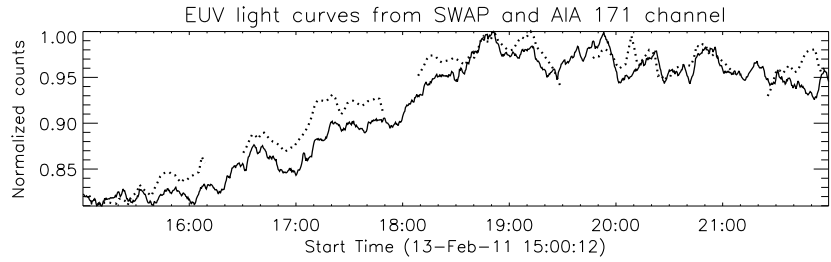}}
\caption{Comparison of integrated EUV emission from SWAP 174~\AA\ (dotted line) and AIA 171~\AA\ (solid line) channels for BP1. The gaps in SWAP data are due to the seasonal eclipses of PROBA2. Despite these gaps, there is good correlation with AIA 171 channel.}
 \label{F-swapaiacor}
 \end{figure}

 \begin{figure}    
 \centering
   \includegraphics[width=0.75\textwidth,clip=]{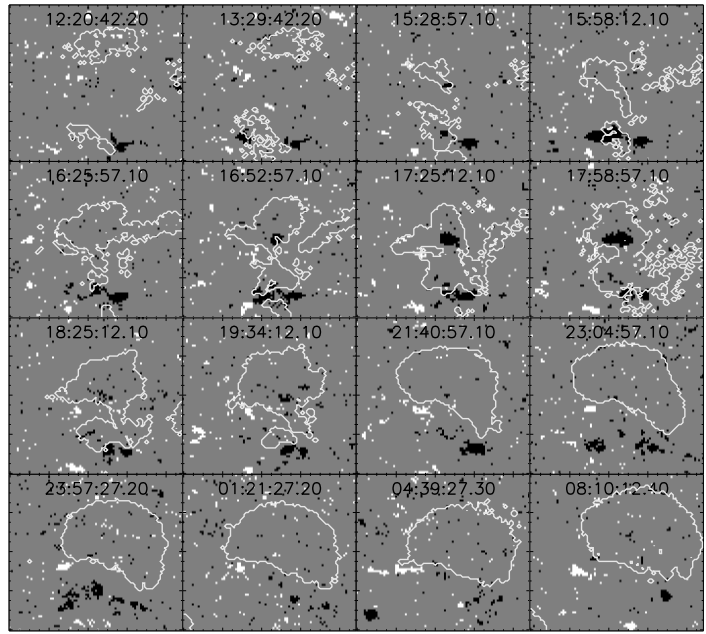}
\caption{Synthetic images showing the positive (white) and negative (black) flux regions above $\pm$25 Gauss. Region in grey represents the background below this threshold. In each image, overplotted contours represent the EUV emission region 1.2 times brighter than the mean value taken in the AIA 193~\AA\ channel.}
 \label{F-binimgs}
 \end{figure}

 \begin{figure}    
 \centering
   \includegraphics[width=0.48\textwidth, clip=]{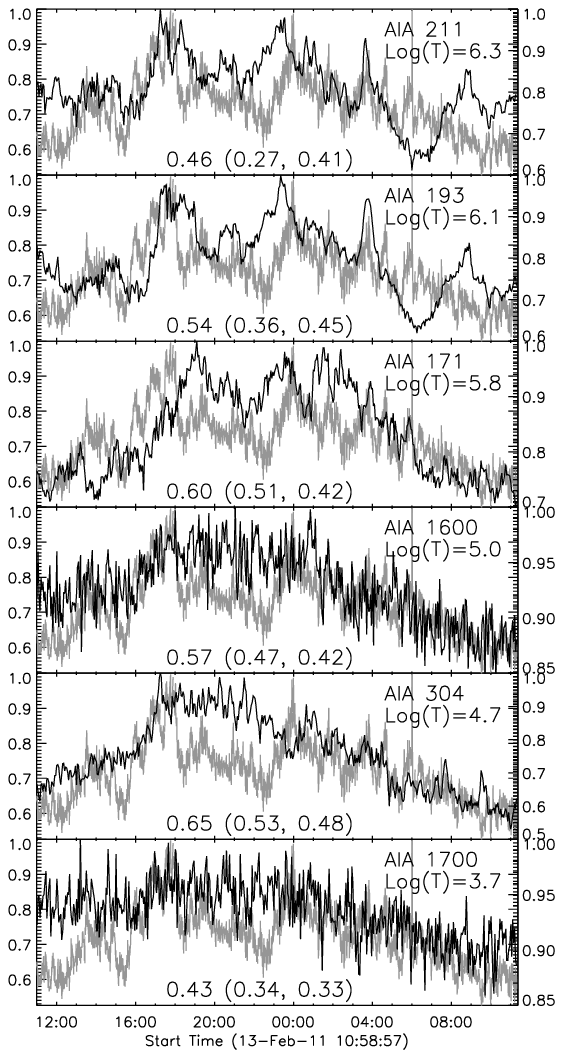}
   \includegraphics[width=0.48\textwidth, clip=]{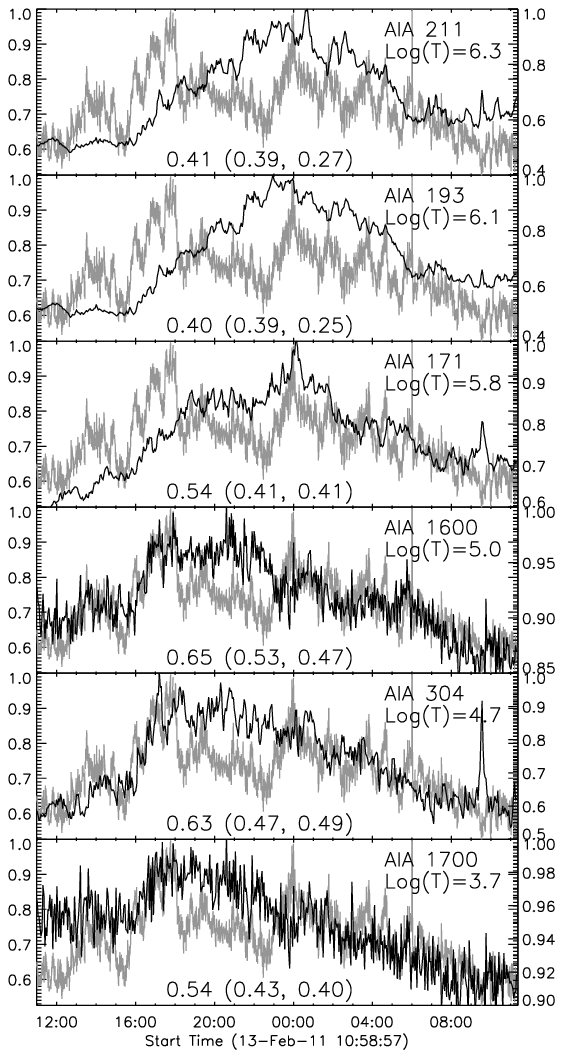}\\(a) \hspace*{2.0in} (b)
\caption{(a) Normalized light curves for BP1 in six different AIA channels, representing different temperature regions above photosphere. These were constructed from the total UV\todash EUV emission brighter than the mean value, in the boxed region that covers this bright point. Overplotted in grey, in each panel, is the total unsigned photospheric flux from the corresponding region above a threshold of $\pm$20 Gauss. Scale for magnetic flux is on left and that for UV\todash EUV emission is on right. All the light curves are normalized to their respective maxima. Characteristic temperatures of individual channels are also listed in each panel. Values listed at the bottom of each panel, are coefficients of correlation between the unsigned flux and the respective intensities. Similar values in the brackets represent these coefficients when the fluxes from individual polarities are compared with the emission.(b) Same as (a) but the UV\todash EUV emission plotted is emission per pixel.}
 \label{F-lcplot}
 \end{figure}

\begin{figure} 
\centering
\includegraphics[angle=90,width=1.0\textwidth]{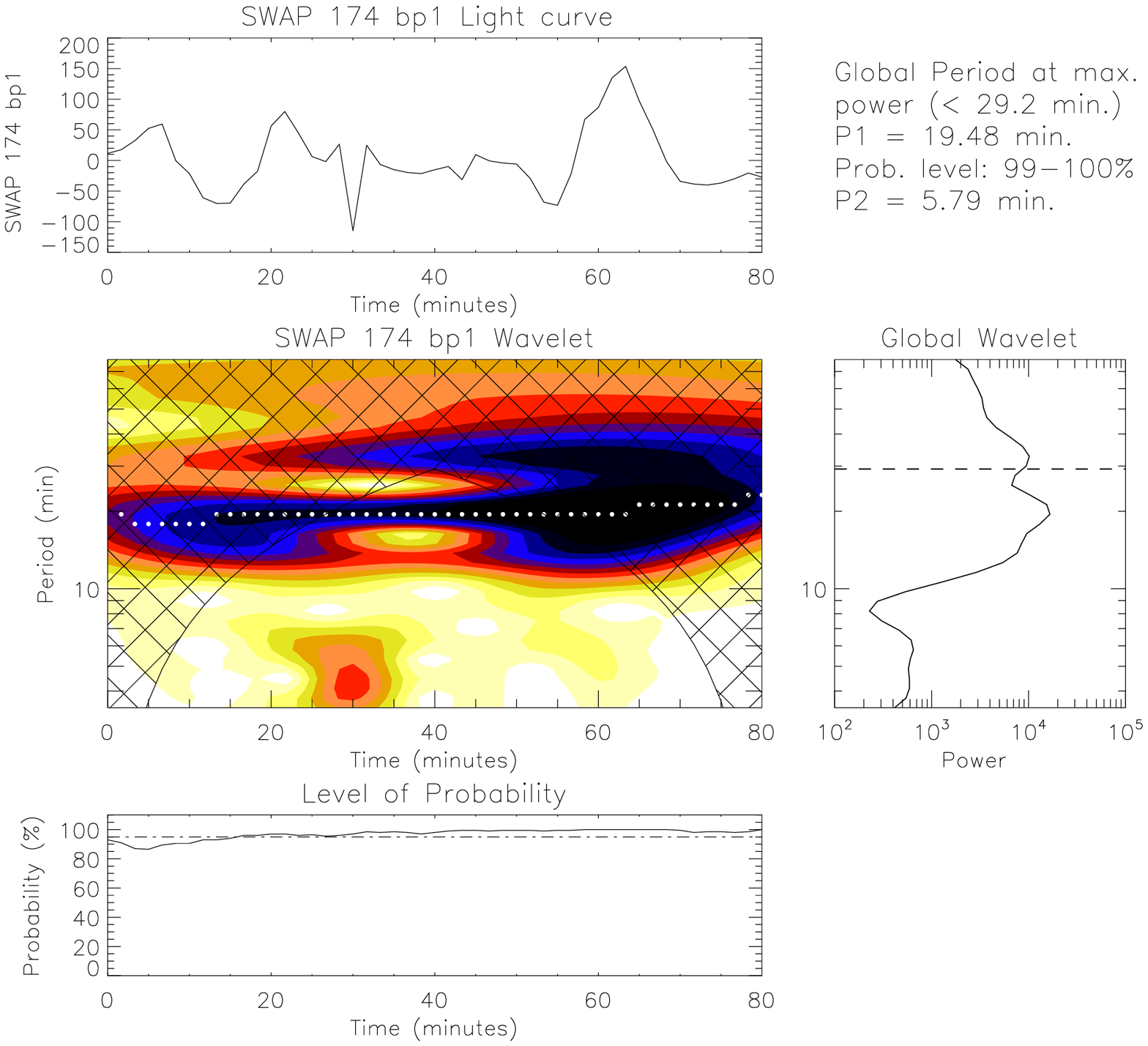} \\
\includegraphics[angle=90,width=1.0\textwidth]{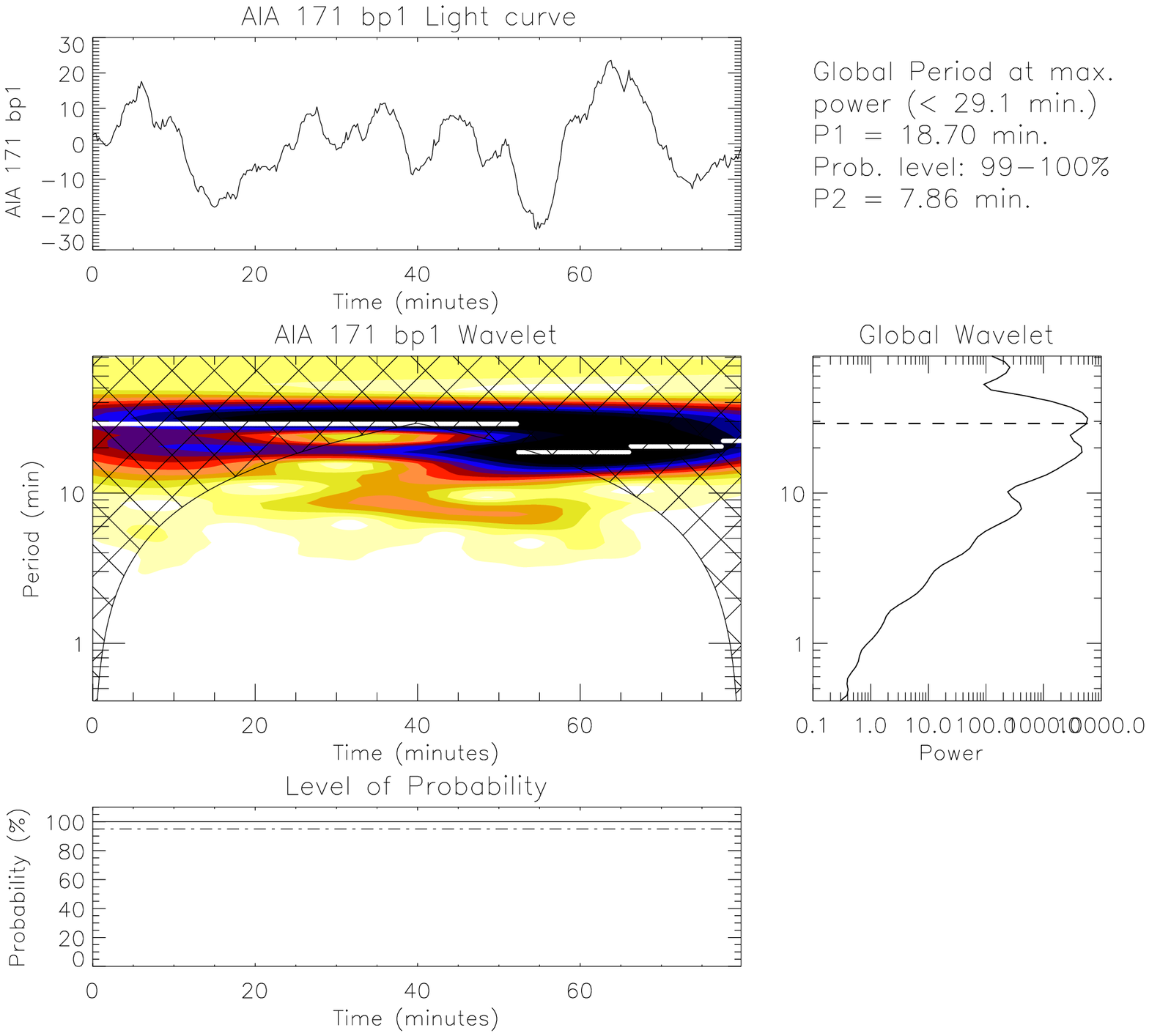} \\
 \caption{Wavelet transformations for BP1 images from SWAP and different channels of AIA (as labeled).  In each set the top panels show the relative (background trend removed) intensity, the central panels show the color inverted wavelet power spectrum, the bottom panels show the variation of the probability estimate associated with the maximum power in the wavelet power spectrum (marked with white lines), and the right middle panels show the global (averaged over time) wavelet power spectrum. The period, measured from the maximum power from the global wavelet, together with probability estimate, is printed above the global wavelet.} 
  \label{F-panel}
\end{figure}

\begin{figure} 
\ContinuedFloat
\centering
\includegraphics[angle=90,width=1.0\columnwidth]{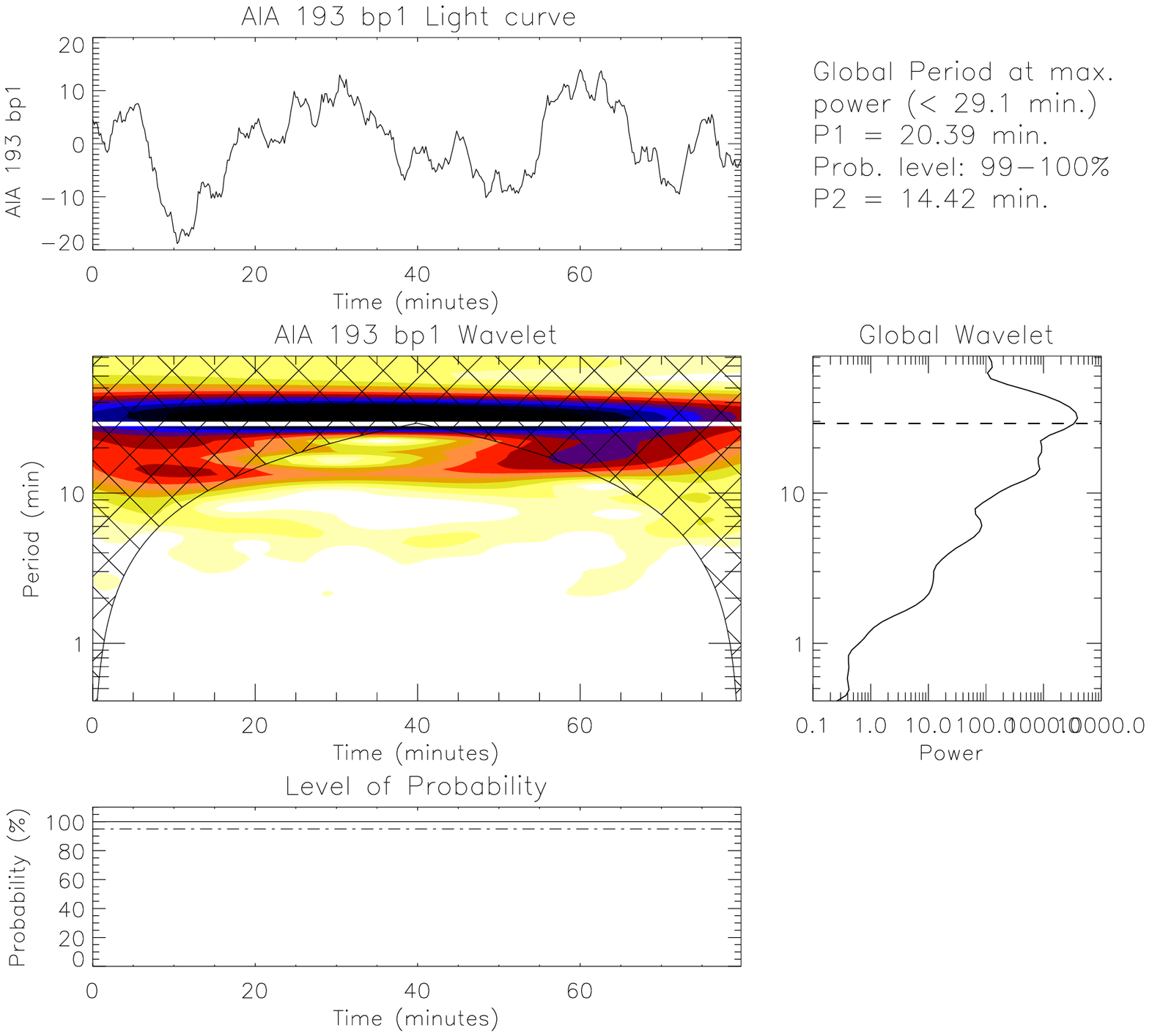} \\
\includegraphics[angle=90,width=1.0\textwidth]{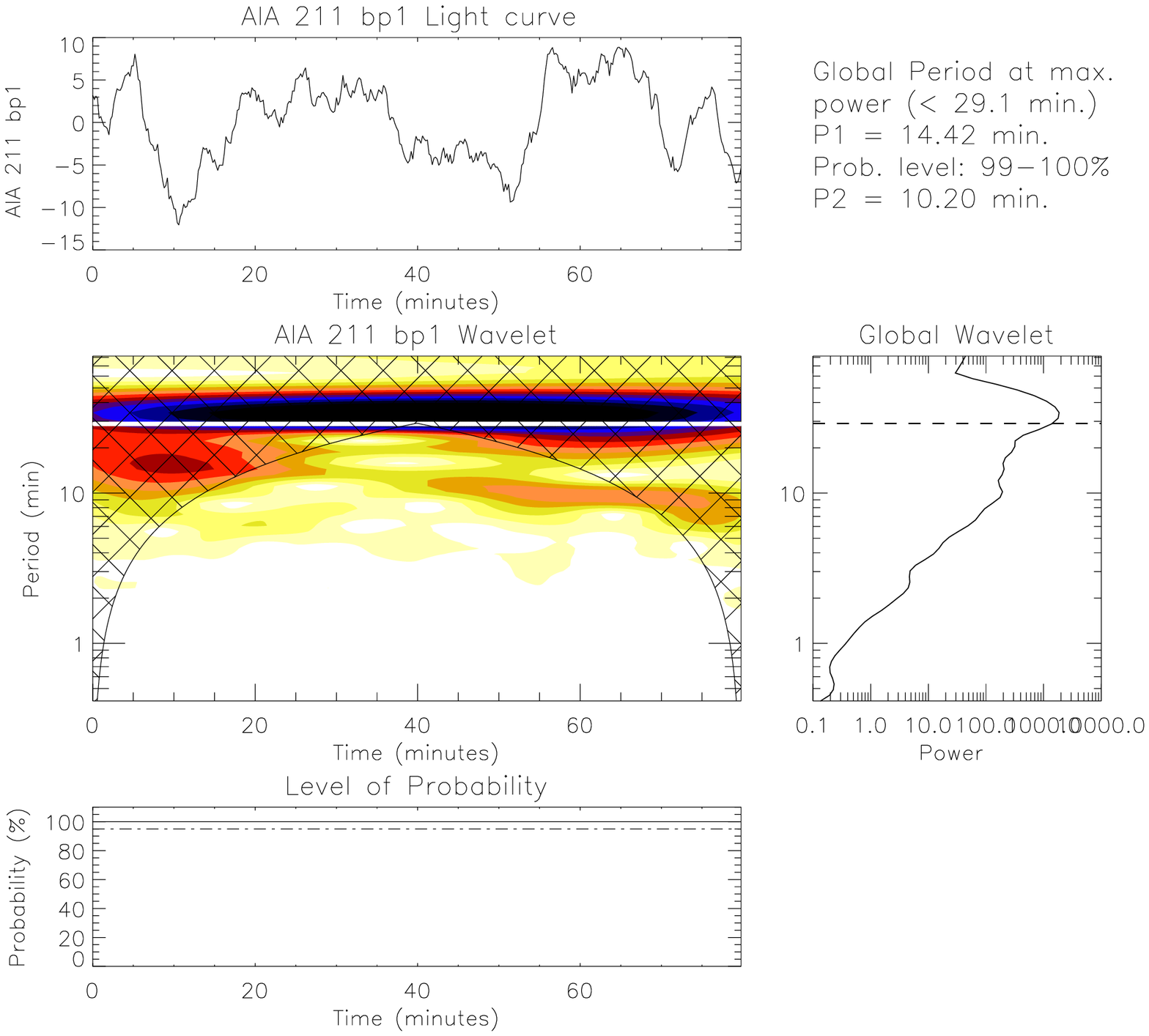}  \\
\caption{Continued.}
\end{figure}

\subsection{Correlation Analysis}
In this subsection, we present the observed relation between magnetic flux and UV\todash EUV intensities of a bright point, marked as BP1 in Figure~\ref{F-swapbps}. We follow this bright point over the course of about 24~hours, starting from 13 February 2011 10:58 UT, in several UV and EUV channels of SDO/AIA. This period covers a part of this bright point's lifetime during its formation. A boxed region of roughly 50~arcsec~$\times$~50~arcsec in size is considered to cover the bright point and it is tracked throughout the observation period in all channels. In this sub-region, we studied the response of UV and EUV intensities in several channels representing different regions above photosphere, to the variations in photospheric magnetic flux obtained from HMI magnetograms. During these observations there was a M6.6 flare, peaking at 17:30 UT. AIA entered its flare mode during this period, so every other frame was obtained with a considerably shortened integration time. This results in some dramatic dips in intensity in our images, where values approach zero in several frames. Though normalizing these observations with respect to exposure time settled most of these cases, there were a few severely underexposed that were replaced by interpolated data. Data from AIA is taken at three minutes cadence. HMI magnetograms are taken regularly with no gaps and hence used directly. \\

For proper comparison of the BP's EUV emission with the corresponding photospheric magnetic flux, we need to eliminate the contribution from background. We set a threshold of $\pm$20~Gauss to estimate the total flux in the box region covering the bright point. We define the total flux, as the sum of all flux contributions of individual pixels above this threshold in the region in each individual snapshot. To obtain the total EUV emission from the bright-point region, we sum over the counts from individual pixels brighter than the mean value of each image. Figure~\ref{F-binimgs} displays synthetic images constructed for several time frames during our observation. In each image, dark and bright pixels correspond, respectively, to the pixels of negative and positive polarities with field strength above $\pm$25~Gauss, while the pixels in grey represent the rest below the threshold. Overplotted contours enclose the bright EUV emission region 1.2 times brighter than the mean value from the nearest snapshot taken in the 193~\AA\ channel of AIA. Slightly higher threshold values were used in constructing these images than in our quantitative analysis to avoid the clutter due to too many points. Nevertheless, these regions represent a vital portion of that used in comparison. It can be seen from this figure that BP1 was in its formation stage during this time and it evolves from a complex structure with multiple changing connectivities to a simple loop system. \\

Figure~\ref{F-lcplot}a shows the normalized light curves of total emission from BP1 in different UV and EUV channels along with total unsigned flux above the set thresholds, from the chosen box region. Normalized emission from each channel is plotted in black and the total unsigned flux is overplotted in grey on each of them. Scale for the magnetic flux is on the left and that for UV\todash EUV emission is on the right of individual panels. Characteristic temperatures of different channels are written in the respective panels. As the values indicate, these channels represent different regions of the solar atmosphere starting from that just above the photosphere (1700~\AA\ channel) to that well inside the corona (211~\AA\ channel). All of the light curves imply a clear good correlation between the photospheric magnetic flux and UV\todash EUV emission. The correlation coefficients between the unsigned flux and the emission from different channels are listed at the bottom of respective panels. Similar values when the fluxes from individual polarities are compared with emission in each of the channels, are listed in brackets, respectively for positive and negative fluxes, separated by a comma. These values indicate that the emission from different channels of AIA correlates well with the unsigned flux better than that with the fluxes from individual polarities. It may be noted that the temporal resolution of HMI (45~seconds) has been degraded to that of AIA (three minutes) to perform this calculation. We also compare the total magnetic flux with the emission per pixel from BP1, which is shown in Figure~\ref{F-lcplot}b. This does not correlate well in the high-temperature channels particularly in the initial stages of the bright point.\\

We also used synoptic data from SWAP in this study. This dataset contains a few larger gaps of nearly 20 minute duration, the result of PROBA2's seasonal eclipses, along with brief, regular gaps due to Large Angle Rotations of the spacecraft. Although PROBA2 follows a dawn\todash dusk Sun-synchronous orbit that provides a continuous view of the Sun for much of the year, from November\todash January PROBA2 experiences occultations of the Sun by the Earth and its atmosphere that can create observational gaps of some tens of minutes per orbit. Larger gaps caused by these seasonal eclipses were left blank and the smaller gaps were filled using linear interpolation. Also data from SWAP are not available from the beginning epoch of this study. For these reasons, instead of directly using the SWAP data in comparison with photospheric magnetic flux, we compare a part of it with corresponding data from 171~\AA\ channel of AIA. Figure~\ref{F-swapaiacor} shows the comparison of these light curves from SWAP 174~\AA\ and AIA 171~\AA\ channels. These are generated by integrating the pixel counts over the chosen box regions and normalizing to their respective maxima. It can be seen that they both agree well despite gaps and interpolation in SWAP data. This indicates that the data from SWAP also may show good correlation with magnetic flux.

\begin{table}[h]
\caption{Bright-point oscillation periods as recorded from SWAP and different AIA channels}
\label{T-oscillations}
\begin{tabular}{ccccccccc}
\hline 
Inst    &    \multicolumn{2}{c}{SWAP}                    & \multicolumn{6}{c}{AIA} \\ 

channel & \multicolumn{2}{c }{174}   & \multicolumn{2}{c }{171 } & \multicolumn{2}{c }{193 } & \multicolumn{2}{c }{211 } \\

Period  & P1          & P2       & P1       & P2     & P1       & P2       & P1     & P2  \\
        &min          & min      &   min   & min     &  min     & min      & min    & min  \\
\hline 
bp1     &  {19.5}       & {5.8}      & {18.7 }   & {7.9 }    &{20.4 }     & {14.4 }     &{14.4}    &{10.2}  \\
  
bp2     &             &          & {20.4}    & {13.2}    & {18.7}     & {4.7}      & {18.7}   & {10.2} \\
  
bp3     &{ 17.9  }      & {8.9}      & {20.4}    & {9.4}     & {26.5}     & {6.6}      & {15.7}   & {24.2} \\
  
bp4     &             &          & {18.7}    & {8.6}     & {20.4}     & {9.4}      & {20.4}   & {9.4} \\
  
bp5     &  	 {19.5} &   {13.8}   & {20.4}    & {7.9}     & {22.2}     & {6.0 }     & {22.2}   & {4.7} \\
  
bp6     &  	 {23.2} &    {8.9}   & {22.2}    & {8.6}     & {22.2}     & {7.2}      & {22.2}   & {6.6} \\
  
bp7     &	 {23.2} & {8.2 }     & {20.4}    & {11.1}    & {17.1}     & {2.1 }     & {18.7}   & {8.6} \\
  
bp8     &       { 25.3} &  { 19.5}   &{ 18.7}    & {6.6}     & {18.7}     & {26.5}     & {18.7}  &  {26.5} \\
  
bp9     &      {  25.3} &{ 10.6}     & {24.3}    & {4.3}     & {22.2}     & {11.1}     & {12.1}   & {0.5} \\
  
bp10 	& 	 {13.8} & {19.5}     &{ 18.7}    & {9.4}     &{ 20.4}     & {7.9 }     & {20.4}   &{ 8.6 }\\
\hline
\end{tabular}
\end{table}
\begin{table}[h]
\caption{Periodicities of BP1 observed for times from February 13, 16:00 UT to 17:00 UT
(T1) and from February 14, 01:UT to 02:00 UT (T2)}
\label{T-ratios}
\begin{tabular}{ccccccc}
\hline 
AIA channels & \multicolumn{2}{c}{171} & \multicolumn{2}{c}{193} &
\multicolumn{2}{c}{211}\tabularnewline

 & T1 & T2 & T1 & T2 & T1 & T2\tabularnewline
\hline
 
P1(min) & {13.2} & {11.12} & {18.7} & {17.15} & {18.7} & {15.72}  \tabularnewline
 
P2(min) & {6.6} & {15.72} & {6.06} & {10.20} & {6.06} &{ 9.35} \tabularnewline
 
(P1/P2) & {2.0} & {0.71} & {3.08} & {1.68} & {3.08} & {1.68}\tabularnewline
\hline 
\end{tabular}
\\
T1 corresponds to time February 13, 16:00 UT to 17:00 UT  \\
T2 corresponds to time February 14, 01:00 UT to 02:00 UT
\end{table}

\subsection{Periodic Variations in Intensity of Bright points}

In this subsection we will discuss the dynamics of the bright points and look for periodic variations in BP intensity. For time-series analysis, it is important to have continuous data with minimal data gaps. Because of the different data gaps in different channels we had to choose the best overlap data between AIA and SWAP. Here we selected an 80-minute time series where these gaps are relatively few. This sequence starts at 19:47~UT on 13 February 2011. A few frames of the SWAP data must be dropped as they are smeared and unusable due to large angle rotations of the spacecraft. These are replaced by linear interpolation. All of the images in the data cube are aligned to the first image using intensity cross-correlation. Sub-field regions of interest (a boxed region of roughly 50~arcsec $\times$ 50~arcsec in size is considered to cover the bright point), shown in Figure~\ref{F-swapbps2}, are then taken both from SWAP and several channels of AIA in order to construct the light curves from integrated intensities. Selection of these BPs is made from SWAP. Using SWAP it is easy to identify the bright points on the solar disk. In low resolution one is able to identify the loop-top (which is very bright) that has very good contrast with the loop foot points.  However, when one looks at the high-resolution images the loop appears to be more uniform and using conventional methods of identifying bright points we would be mistaken if we use only the high-resolution data. 

We used wavelet analysis to study the oscillations in these bright points. Details of the wavelet analysis, which provides information on the temporal variation of a signal, are contained in \inlinecite{1998BAMS...79...61T}. For the convolution with the time series in the wavelet transform, we chose the Morlet function, a complex sine wave modulated by a Gaussian that is discussed in detail in \inlinecite{1998BAMS...79...61T}. The periodicities shown in Table~\ref{T-oscillations} had their background trend removed by subtracting from the original time series an approximately 150-point ($\approx30$~minutes) and 18-point ($\approx30$~minutes) running average respectively from the AIA and SWAP light curves.

In the resulting plots of the wavelet spectrum,  the cross-hatched regions are locations where estimates of the oscillation period become unreliable. This is the so-called cone-of-influence (COI) as described by \inlinecite{1998BAMS...79...61T}. As a result of the COI, the maximum measurable period is shown by a dashed line in the global wavelet-spectrum plots.  The probability estimate was calculated using the randomization method with 200 permutations as outlined in detail by  \inlinecite{2001A&A...380L..39B}. Below the wavelet-power spectrum, in the lowest panels, we show the variation of the probability estimate, calculated using the randomization technique, associated with the maximum power at each time in the wavelet-power spectrum. The location of the maximum power is indicated by the over-plotted white lines. Above the global wavelet spectrum we have displayed the period of the peak power measured in the global wavelet spectrum. Only those oscillations with a probability greater than 95\% are considered to be significant. 

In Figure~\ref{F-panel} we show oscillations observed for BP1. The observed periodicities corresponding to all ten bright points that we studied are tabulated in Table~\ref{T-oscillations}. In Table~\ref{T-oscillations}, P1 refers  to the most significant peak and P2 refers to the second most significant peak in the global wavelet spectrum with  confidence level higher than 95\%.  We observe periodicities of the BPs are in the range of 10\todash 25 minutes (P1). There are reports of long-period oscillations in BPs \cite{2008A&A...489..741T} as well as reports of around five-minute  oscillations as seen in EIS observations \cite{2010MNRAS.405.2317S}. 

The period ratio [P1/P2] of acoustic oscillations can be used as a tracer of density stratification in the large-scale coronal loops \cite{2006A&A...448..763M,2009SSRv..149....3A,2010A&A...515A..41M} and we do find some ratios of P1/P2 that might be useful for coronal seismology. It has already been pointed out that in the BPs, the intensity oscillations may be generated due to recurrent magnetic reconnection \cite{2004A&A...418..313U,2006A&A...446..327D,2008A&A...489..741T}. \inlinecite{2011MNRAS.415.1419K} have recently observed different P1/P2 ratios for different BPs. Accordingly, depending on whether [P1/P2] is greater or less than two, they divided BPs in two categories. For BPs in which $\textrm{P1/P2}<2$ density stratification is the most probable explanation, while others BPs, for which $\textrm{P1/P2}>2$, magnetic-field divergence is the most probable explanation for the observed ratios \cite{2008A&A...486.1015V}. 

We made an attempt to find period ratios for BP1 for two different time intervals: 13 February 16:00~UT to 17:00~UT (referred to as T1) and 14 February 01:00~UT to 02:00 UT~(referred to as T2), in AIA 171, 193, and 211 channels. Periods found in the time intervals selected and their ratios are tabulated in Table~\ref{T-ratios}. Time interval T1 is the period where changes in magnetic flux and connectivities are most prominent (since there are apparent changes in field  we assume this is due to repeated reconnection), whereas during time interval T2 changes are not abrupt. In Table ~\ref{T-ratios} we see that in the AIA 193 and 211 channels $\textrm{P1/P2}>2$ during the time interval T1, while it is $\textrm{P1/P2}<2$ during T2. This suggests both that abrupt or large changes in magnetic flux have direct effect on the oscillation periodicities and that magnetic activity, such as field divergence,  may come in to play in such situations. Also note that most of the theoretical work on P1/P2 ratio concerns actual eigenmodes of the system. If the oscillations are driven by repeated reconnection, it is not immediately obvious the system is oscillating with its eigenfrequency. The periods could be imposed by the timescale on which the reconnection is occurring. In the latter case, the theory of the P1/P2 ratio might not necessarily apply.


\section{Discussions}   
    \label{S-conclusions}

Bright points are found to be always associated with opposite-polarity poles in photospheric magnetograms \cite{1976SoPh...49...79G}. Nevertheless, we see a lot of activity in HMI magnetograms corresponding to the location of BP1. We see multiple connectivities in BP1. From the formation stage of the bright point, and throughout its entire lifetime, there is continuous emergence of new magnetic flux and there is good overall correlation between the total magnetic flux and emission from the bright point in several UV and EUV channels. This suggests that the emerging flux interacts with the existing fields, which results in reconnection at the site of bright point. This, we conjuncture, leads to a series of continuous periodic brightenings as seen in different EUV lines. Within the small region around BP1's location, we see at least three major connectivities in the formation stage.  We also observe that, in high-resolution images, the bright point looks like a miniature active region with multiple connectivities and with several loop structures.

\inlinecite{2001ApJ...547.1100H} studied quiet-Sun loops, from EIT and MDI data. They have given a threshold of $3 \times 10^{18} $~Mx of magnetic flux to be associated with brightening in Fe~{\sc xii} corona. This number depends on the loop foot-point distance we see in EUV lines and the distance of the corresponding bipoles in the photosphere. In the case of bright points, which are small loops compared to typical quiet-Sun loops, this number differs. It also depends on which of the EUV lines we see brightenings in. In the case of BP1 we see brightenings in different channels of AIA and SWAP corresponding to a few times $10^{18}$~Mx. This supports the proposition discussed by \inlinecite{2003A&A...398..775M} that appearance of brightening at a certain temperature depends strongly on the magnetic energy released and therefore, on the magnetic-field strength. Combining our observational results for BP1 with bright-point observations by \inlinecite{2003A&A...398..775M} and observations of quiet coronal loops by \inlinecite{2001ApJ...547.1100H} one can conclude that size of the loops is also important as far as the brightening is concerned. 

Several authors \cite{1999ApJ...510L..73P,2001ApJ...547.1100H,2004A&A...418..313U,2008A&A...492..575P} reported a positive correlation between the EUV and X-ray emission with the underlying photospheric magnetic flux. \inlinecite{1999ApJ...510L..73P} observed that the photospheric magnetic flux is highly temporally correlated with X-ray and EUV emissions. \inlinecite{2004A&A...418..313U} found that the unsigned magnetic flux correlates well not only with the total EIT flux but also with EIT flux per pixel. They suggested that, as the magnetic flux becomes stronger, coronal emission from the BP increases not just because the emission area increases, but also because the intrinsic emission of the BP itself increases. For BP1, although we studied only a part of its lifetime, there exists a good correlation between the total magnetic flux and the total emission in different temperature channels (Figure~\ref{F-lcplot}a). But when the average emission per pixel is considered (Figure~\ref{F-lcplot}b), the correlation is not so good in the high-temperature channels, particularly during the initial phase of the bright point. However, the correlation is good for the entire duration in the channels that represent transition-region temperatures or below. So, the relationship between the intrinsic emission of the BP with increasing in magnetic flux indeed requires further confirmation.

Nevertheless, the good correlation observed between total emission from the BP in six different channels, representing different temperature layers above the photosphere, indicates strongly that this BP is fed by magnetic reconnection; supporting the prevailing idea. \inlinecite{2008A&A...492..575P} have studied the structure and dynamics of a bright point from \textit{Hinode}, SOHO, and TRACE data. They found a positive correlation between weaker magnetic flux associated with one of the footpoints in a bipole and the total XRT brightening. In the case of BP1, negative flux is observed to be much stronger than the positive flux, but the emission is correlated well with the total unsigned flux, rather than with the fluxes from individual polarities.

This result will have implications for both coronal heating and filling factor. In particular, we must point out here that we have selected only a few bright points to study the dynamical properties of CBPs; in particular we have focused our attention on one representative BP1 to show the general behavior of the BPs. A detailed statistical work will be essential in order to address whether or not the CBPs will be able to supply substantial energy flux to the energy budget of coronal heating.

Finally we attempted to study whether or not the periodic brightenings of the CBPs can be used for coronal seismology. The observed periods are generally longer than those observed previously (10\todash 25~minutes, See Tables~1 and 2). There are some indications that the period ratio P1/P2 changes during the lifetime of the bright point, which may suggest a scenario of either magnetic-field divergence or density stratification as claimed by \inlinecite{2011MNRAS.415.1419K}. But we also must point out that we have not yet found a convincing theory that the period ratio can indeed be used as a tool for inferring the driver of such periodic brightenings. However, a good correlation between magnetic flux associated with the foot points and the intensity brightenings provides support for a repeated reconnection scenario. Further statistical work on this is required to come to a conclusion, which we intend to do in the future.

\begin{acks}
We thank the referee for their valuable comments. The AIA data used here is courtesy of SDO (NASA) and AIA consortium. DB and SKP thank the Guest Investigator Program of the PROBA2 mission, which supported this study. We would also like to thank the SWAP team, in particular D. Berghmans and Anik de Groof for their help at different stages of reduction of SWAP data. SWAP is a project of the Centre Spatial de Liege and the Royal Observatory of Belgium funded by the Belgian Federal Science Policy Office (BELSPO).
\end{acks}

\bibliographystyle{spr-mp-sola.bst}
\bibliography{chandu.bib}
\end{article} 

\end{document}